\documentclass[11pt,DIV=calc]{scrartcl}

\usepackage{amssymb}
\usepackage{amsmath}
\usepackage{amsfonts}
\usepackage{bm}
\usepackage{authblk}

\usepackage[utf8]{inputenc}
\usepackage[T1]{fontenc}
\usepackage{lmodern}

\usepackage{scrlayer-scrpage}
\usepackage[shortlabels]{enumitem}

\usepackage{booktabs}

\usepackage{array}

\usepackage[usenames,dvipsnames,svgnames,table]{xcolor}

\usepackage{caption}
\usepackage{subcaption}

\usepackage[british]{babel}

\usepackage[all,error]{onlyamsmath}
\usepackage{mathtools}
\usepackage{amssymb}
\usepackage{amsthm}
\usepackage{comment}

\usepackage{hyperref}
\hypersetup{hidelinks}

\usepackage{siunitx}
\usepackage{longtable}
\usepackage[indexonlyfirst,nogroupskip,smallcaps,acronym]{glossaries}
\usepackage{float}
\usepackage{csquotes}
\MakeAutoQuote{«}{»}

\usepackage[doi=false,url=false,giveninits=true,hyperref,maxbibnames=10,backend=biber]{biblatex}
\bibliography{paper}

\usepackage{graphicx}

\usepackage{tikz}
\usetikzlibrary{shapes.geometric, arrows,backgrounds}

\usetikzlibrary{intersections,babel,positioning,patterns}
\usepackage{pgfplots}
\pgfplotsset{compat=1.14}
\usepgfplotslibrary{fillbetween}

\usepackage{url}

\usepackage{float}
\usepackage[textsize=tiny]{todonotes}

\newtheorem{theorem}{Theorem}

\theoremstyle{definition}

\usepackage[dvipsnames]{xcolor}

\usepackage[retainorgcmds]{IEEEtrantools}
\usepackage{caption}

\newtheorem{example}[theorem]{Example}

\newcommand{\E}{\mathbb E}

\def\P{{\mathbb P}}

\makeatletter
\@addtoreset{equation}{section}

\makeatother

\recalctypearea

\newglossaryentry{targetvelocity}
{
  name={\ensuremath{v}},
  description={is the (constant) velocity maintained by the driver},
  sort=v
}

\newglossaryentry{frequency}
{
  name={\ensuremath{f}},
  description={is the frequency of object detection and depth
    estimation system updates},
  sort=f
}

\newglossaryentry{gravitation}
{
  name={\ensuremath{g}},
  description={is the gravitational constant},
  sort=g
}

\newglossaryentry{threshold}
{
  name={\ensuremath{c}},
  description={is the threshold distance to objects at which automatic brakes
    engage},
  sort=c
}

\newglossaryentry{friction}
{
  name={\ensuremath{\mu}},
  description={is the known and fixed surface friction of the road},
  sort=mu
}

\newglossaryentry{brakedist}
{
  name={\ensuremath{b}},
  description={is the braking distance of the vehicle when travelling at the
    target velocity},
  sort=b
}

\newglossaryentry{bufferdist}
{
  name={\ensuremath{r}},
  description={is the buffer distance},
  sort=db
}

\newglossaryentry{numestimates}
{
  name={\ensuremath{k}},
  description={is the number of updates of the object detection and depth
    estimation system performed from when an object enters the threshold
    distance to when it enters the braking distance},
  sort=m
}

\newglossaryentry{acceldist}
{
  name={\ensuremath{d_{a}}},
  description={is distance driven while accelerating from standstill to the
    target velocity},
  sort=da
}

\newglossaryentry{targetlevel}
{
  name={\ensuremath{\varepsilon}},
  description={is the target level on the expected number of collisions per
    kilometre},
  sort=epsilon
}

\newglossaryentry{position}
{
  name={\ensuremath{X_{t}}},
  text={\ensuremath{X}},
  description={is the position of the vehicle (at time \(t\))},
  sort=Xt
}

\newglossaryentry{velocity}
{
  name={\ensuremath{V_{t}}},
  text={\ensuremath{V}},
  description={is the velocity of the vehicle (at time \(t\))},
  sort=Vt
}

\newglossaryentry{truedist}
{
  name={\ensuremath{L_{t}}},
  text={\ensuremath{L}},
  description={is the true distance to the closest object in front of the
    vehicle (at time \(t\))},
  sort=Lt
}

\newglossaryentry{estdist}
{
  name={\ensuremath{D_{t}}},
  text={\ensuremath{D}},
  description={is the estimated distance to the closest object in front of the
    vehicle, given by the depth estimation system (at time \(t\))},
  sort=Dt
}

\newglossaryentry{pedestrians}
{
  name={\ensuremath{N}},
  description={is the random measure describing locations of stationary objects
    (pedestrians) on the road},
  sort=N
}

\newglossaryentry{collisions}
{
  name={\ensuremath{\mathcal{T}}},
  description={is the set of collisions},
  sort=T
}

\newglossaryentry{numcollisions}
{
  name={\ensuremath{N_{c}}},
  description={is the number of collisions},
  sort=Nc
}

\newglossaryentry{numpedestrians}
{
  name={\ensuremath{N_{o}}},
  description={is the number of obstacles during the drive},
  sort=No
}

\newglossaryentry{hitvelocity}
{
  name={\ensuremath{V_{h}}},
  description={is the velocity at which the object under consideration is hit},
  sort=Vh
}

\newglossaryentry{drivedist}
{
  name={\ensuremath{K}},
  description={is the total distance to drive within the driving session},
  sort=K
}

\newglossaryentry{esterror}
{
  name={\ensuremath{Z^{1}, \dotsc}},
  text={\ensuremath{Z}},
  description={are the estimation errors while vehicle approaching a single obstacle},
  sort=Vh
}

\newacronym{odd}{odd}{Operational Design Domain}

\newacronym{sotif}{sotif}{Safety Of The Intended Functionality}

\title{Self-driving car safety quantification via component-level analysis
}

\author[1]{Juozas Vaicenavicius}
\author[1,2]{Tilo Wiklund}
\author[1]{Austė Grigaitė}
\author[1]{Antanas Kalkauskas}
\author[1]{Ignas Vysniauskas}
\author[1]{Steven Keen}
\affil[1]{Sensmetry UAB, Lithuania}
\affil[2]{Department of Mathematics, Uppsala University, Sweden}

\date{}
\begin{document}

\maketitle

\begin{abstract}
  In this paper, we present a rigorous modular statistical approach for arguing safety or its insufficiency of an autonomous vehicle through a concrete illustrative example. The methodology relies on making appropriate quantitative studies of the performance of constituent components. We explain the importance of sufficient and necessary conditions at the component level for the overall safety of the vehicle as well as the cost-saving benefits of the approach. A simple concrete automated braking example studied illustrates how separate perception system and operational design domain statistical analyses can be used to prove or disprove safety at the vehicle level. 
\end{abstract}


\section{Introduction}

Road vehicles can cause immense harm if driven inadequately and autonomous vehicles are no exception. As is customary in engineering domains concerned with developing systems that have a potential to cause harm, the safety of autonomous driving systems is also being addressed by international standards and regulations, compliance with which assures the public about the absence of unacceptable level of risk. 

There are multiple standards existent and in development that are relevant to AD, such as ASAM OpenX standard family \cite{opendrive, openscenario, opencrg, openodd}, and addressing AD safety, such as UL 4600 \cite{ul4600}, among which the most prominent being ISO21448 'Safety of the Intended Functionality (SOTIF)' \cite{iso_pas_2019} and the ISO26262 standard series for Functional Safety (FS) \cite{iso201126262} and their updates while the taxonomy and definitions are provided in J3016 \cite{j3016}. The development of the SOTIF standard follows in the footsteps of by now well-established and ubiquitous ISO26262 functional safety standard and is intended to address the safety aspects of autonomous driving functions that are not addressed by ISO26262. It is a simplistic but still useful analogy to think about SOTIF as governing the safety of the driver (autonomous driving system) and the ISO26262 series governing the safety of the more traditional aspects of electrical and electronic automotive systems that are not much dependent on perceiving and interacting with the complex surrounding environment. What is common in the application of both standards is that careful quantitative assessment of risks must be provided in order to justify that the expected harm during system operation will be below pre-specified acceptable threshold levels.

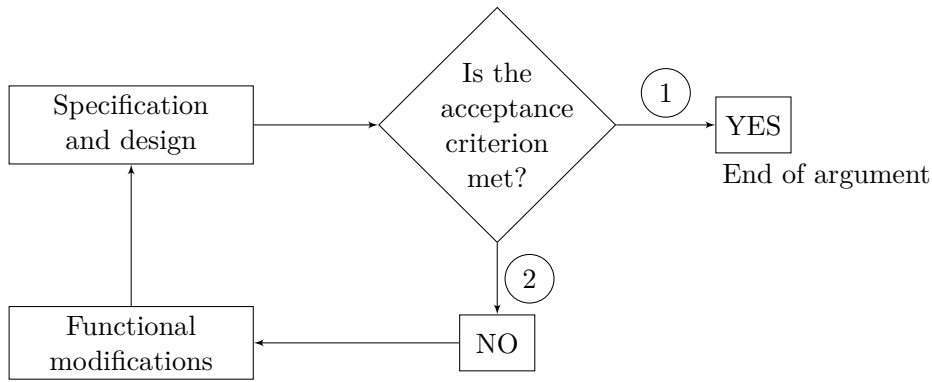
\begin{figure}
\resizebox{\columnwidth}{!}{
\tikzstyle{decision} = [diamond, draw,  
text width=4em, text badly centered, node distance=5cm, inner sep=0pt]
\tikzstyle{block} = [rectangle, draw , 
text width=8em, text centered,  minimum height=2em]
\tikzstyle{answer_block} = [rectangle, draw , 
text width=2em, text centered,  minimum height=2em]
\tikzstyle{line} = [draw, -latex']
\tikzstyle{edgelab} = [circle, draw,  minimum size = 0.5mm]

\begin{tikzpicture}[node distance = 3cm, auto]
\node [block] (sd) {Specification and design};
\node [block, below of=sd] (fm) {Functional modifications};
\node [decision, right of=sd, node distance=5cm] (q) {Is the \\{acceptance} criterion met?};
\node [answer_block, right of=q, xshift = 0.5cm] (yes) {YES};
\node (sub) [below of = yes, yshift = 2.3cm, xshift = 1cm] {End of argument};
\node [answer_block, below of=q] (no) {NO};
\path [line] (fm) -- (sd);
\path [line] (sd) -- (q);
\path [line] (q) --  node[edgelab,anchor = south, yshift= 0.1cm] {1}(yes);
\path [line] (q) --node[edgelab, xshift= 0.1cm] {2}(no);
\path [line] (no) --node[anchor = south, yshift= 0.1cm] {}(fm);
\end{tikzpicture}
}
\caption{The basic iterative development process diagram at the core of the ISO21448 standard (starting at `Specification and design'). This article presents how rigorous quantitative arguments for the process steps 1 and 2 can be obtained in an illustrative example studied. For a detailed ISO21448 diagram, see Figure 9 in \cite{iso_pas_2019}.}
\label{F:IterProcess}
\end{figure}

A common way to quantify safety is in terms of risk, which mathematically is
expressed as an expectation \(\E[H]\) with \(H\) representing a measure of
actual harm (e.g. number of deaths/injuries/etc.) calculated over a unit
distance or time of driving. A quantitative safety target is expressed as a
required upper bound for the risk, i.e.~\(\E\left[H\right]<\;\epsilon\), where
\(\epsilon\) is a small number representing an acceptable level of risk. Thus
the validity of the inequality is considered as the acceptance criterion for the
autonomous vehicle safety (see \cite{iso_pas_2019}, particularly, Annex C). We
note that the risk quantity \(\E[H]\) is not just a theoretical construct but
will manifest itself in reality through collisions, insurance claims, lawsuits,
and callbacks during the vehicle operation, demanding painstaking quantitative
safety assessment. The schematic diagram depicting the basic iterative
development processes at the core of ISO 21448 appears in Figure
\ref{F:IterProcess}.

 Arguing whether an autonomous vehicle meets quantitative safety targets is a complex matter and still a largely open research question of tremendous societal impact. There are three main existing approaches available as potential solutions/methodologies. The first is to estimate the risk via real-world autonomous vehicle testing. Unfortunately, a direct approach is practically infeasible in most situations as the distances required to be driven to obtain results with a desired level statistical accuracy (confidence) are enormous \cite{kalra_driving_2016}. Not to mention that the public is directly exposed to an unknown level of risk exposing a moral issue. Another unfavourable practical aspect of direct validation is the resulting rigidity of the system design as modification of most modules of the AD system in principle requires performing the same excruciatingly lengthy and costly validation again from scratch. The second approach is to perform closed-course testing which has been adopted by, for instance, Waymo, as it is described in section 3 of Waymo Safety Report \cite{Waymo_report}. Unlike real-world testing, this method does not put the public at risk, but the requirement of practically infeasible distance needed to be travelled remains and this also raises a question of whether the closed course is representative of the real world. The third approach is to build an accurate representative model of the autonomous system, the environment, and their interactions, and then use the multitude of simulated scenarios generated by the model to estimate the risks of interest. Simulation for AD testing is a very active research area both in industry and academia with the papers \cite{shah_airsim_2018, rosique_systematic_2019,kelly_scalable_2018, chamberlain_safe_2019, dosovitskiy_carla_2017, rosique_systematic_2019}, a non-exhaustive list providing an overview of the area. With the potential benefits of simulation aside, whether it is possible to build a sufficiently accurate model of all the relevant complex real-world phenomena and justify its real-world equivalency necessary to obtain an accurate real-world risk estimate is unclear \cite{tiwari_simulation_2019}. Though the acceptance criterion is formulated at the vehicle level, another highly desired feature of safety argumentation is support for modular designs so that component-level and Operational Design Domain (ODD) analyses can be combined into a final safety argument at the vehicle-level. The potential benefits of modular safety argumentation are reduced validation costs as separate component-level and ODD analyses can be cheaper and reusable, comparing to vehicle-level random road testing. To this end, in this article we present an example application of a methodology for producing rigorous quantitative arguments for arguing safety or its insufficiency for autonomous driving functions (steps 1 and 2 in Figure \ref{F:IterProcess}) with the following three desirable features; arguments that are:
 
\begin{itemize} \itemsep0em 
\item \textit{modular}, utilising component-level analysis, and
\item \textit{statistically rigorous}, with the uncertainty about meeting the acceptance criterion quantified,
\item \textit{cost-efficient}, through reusable analyses of constituent modules.
\end{itemize}

The fundamental guiding principle in our approach is to avoid making any strong modelling assumptions that are not known for certain or that cannot be justified well statistically from data. Instead we rely on probabilistic worst-case thinking and choose to be conservative about any assumptions that cannot be justified. This typically boils down to deliberately making overly pessimistic or overly optimistic assumptions comparing to reality, depending on the type of argument (for safety or for lack of it) we want to make.  In a sense, the approach attempts to be `robust' or `model-free' as it is consistent with many possible true real-world models.  

In this paper, we conduct an analysis of a simple problem with the aim to illustrate the key ideas and concepts encountered when applying this approach. No claim is made regarding realism of the example or that the maxim of pessimism or optimism is upheld in every detail as such work is outside the scope of this paper. Rather, the analysis presented should be taken as educational to elucidate the main guiding principles in the process or as a potential starting point for developing more detailed analyses of real autonomous driving systems.

The structure of this document is as follows. A description of a simple autonomous driving system, its Operational Design Domain (ODD)
and a formalised problem description can be found in the next section. Afterwards, we proceed to describe the mathematical and statistical methodology for analysing the safety aspects of the perception component as well as the autonomous driving system as a whole (steps 1 and 2 in Figure \ref{F:IterProcess} and the topic of Section \ref{S:args}). Next, a discussion on refining the model and analysis is provided in Section \ref{S:refining}. Finally, we finish some concluding remarks.

\section{Problem description}

Suppose we are an organisation wishing to deploy a new automated driving system responsible for driving at a fixed constant velocity on a straight road and for stopping in front of stationary objects that can occasionally appear on this road. In order to obtain system certification for use within a certain type of environment and mission profile, known as the Operational Design Domain (ODD), we must produce an analysis of its behaviour and based on it provide an argument whether the system is safe or unsafe to deploy in terms of meeting or not meeting a quantitative safety target.

The rest of this section proceeds to formalise this scenario. First, the \acrshort{odd} is specified and a simple model is defined. Next we specify an
abstracted target for the performance (behaviour) of the system in terms of the model set out in the subsection before. Finally a specification of the braking
system, to the extent that its functions can be understood, is given.

\subsection{\acrlong{odd} specification and model}

An autonomous driving function is to be validated for use.  The design goal of the system is to avoid collision events with other ego-actors or structures while driving within the defined \acrshort{odd} described below.  Alternatively, this system can be viewed as a debris detection function with an automated braking functionality as part of a more complex autonomous driving system. It is attached to a vehicle which, apart from accelerating and braking, is driven in such a way as to maintain a constant velocity \gls{targetvelocity} along a straight, homogeneous, single-lane, one-way road. The surface friction \gls{friction} of the road is constant and specified as part of the \acrshort{odd} for which the system is to be certified. Along the road path, stationary objects obstructing the lane randomly appear ahead of the vehicle during operation; the properties and frequencies of these obstructing objects are given only informally in terms of, for example, a geographic region and/or time of day. For the purpose of this document we will restrict the analysis to the main driving session and exclude an analysis of the vehicle starting and stopping at a departure point and terminus.

In order to transform this problem statement from the engineering language above into a mathematical language of suitable rigour to permit probabilistic analysis, the problem space is defined as follows. Let us now describe a single driving session of a \(\gls{drivedist}\,\si{\kilo\metre}\) distance within the \acrshort{odd}. Let \(\gls{position}_{t} \in [0, \gls{drivedist}]\) be the position, along the road, relative to the starting point, of the front part of the vehicle at time \(t \geq 0\). Also let \(\gls{velocity}_{t}\) be the velocity and acceleration of the vehicle at time \(t \geq 0\). All three will be assumed to satisfy standard regularity conditions and that \(\frac{d}{dt}\gls{velocity}_{t} = \frac{d^{2}}{dt^{2}}\gls{position}_{t}\) for each realisation of the vehicle trajectory. To capture the positions of obstacles let \(\gls{pedestrians}_{t}\) be the point processes (random measure) \cite{daley_introduction_2007} such that for any interval \(I \subseteq
[0, \gls{drivedist}]\) the number of obstacles in \(I\) at time \(t\) is \(\gls{pedestrians}_{t}(I)\). The total number of obstacles that will be
encountered is given by \(\gls{numpedestrians} = \gls{pedestrians}_{0}([0,\gls{drivedist}])\). As time goes on the number reduces due to the vehicle
stopping within the specified distance \(r\). Finally let
\begin{equation*}
  \gls{collisions} =
  \{ x \in [0, \gls{drivedist}] : \exists t \; \gls{position}_{t} = x \land \gls{velocity}_{t} > 0, \gls{pedestrians}_{t}(\{x\}) \geq 1 \}
\end{equation*}
be the set of collisions and \(\gls{numcollisions} = |\gls{collisions}|\) the number of collisions.

For completeness, before entering the \(K\,\si{\kilo\metre}\) road segment on which collisions are counted, the vehicle is assumed to have been travelling at speed \gls{targetvelocity} for distance \gls{threshold} without any objects. This is to ensure that the vehicle is not already within a dangerous collision distance at the beginning of the segment under analysis. 

\subsection{Quantitative objective}

The goal of any safety analysis would be to establish that the number of probable collisions over the distance travelled  \(\gls{numcollisions}/\gls{drivedist}\) is sufficiently small for reasonable sizes of \(\gls{drivedist}\). That is to say, we would wish to establish an inequality of the type
\begin{equation}
  \E[\gls{numcollisions}/\gls{drivedist}] \label{eq:acceptance}
  \leq
  \gls{targetlevel},
\end{equation}
where \(\E[\gls{numcollisions}/\gls{drivedist}]\) is the expected number of collisions generated by the system in question within the defined \acrshort{odd} and \(\gls{targetlevel} > 0\) is some pre-specified quantitative safety target level given to us.

\subsection{Autonomous driving system specification}
\label{sec:perception}

The vehicle is equipped with an autonomous driving functionality whose specification is as follows. A perception system for detecting objects in front of the vehicle performs combined object detection \& depth estimation using the available sensors and a perception algorithm for interpreting the sensor data. The perception function runs at frequency \gls{frequency} and provides a new distance estimate to the closest object every \(1/\gls{frequency}\) seconds. The only thing given about the perception system is that it operates on a frame-by-frame basis using only the newest sensor data that was not used previously for earlier distance-to-object estimates.

\begin{figure}[h]
  \centering{}
  \includegraphics[width=\textwidth]{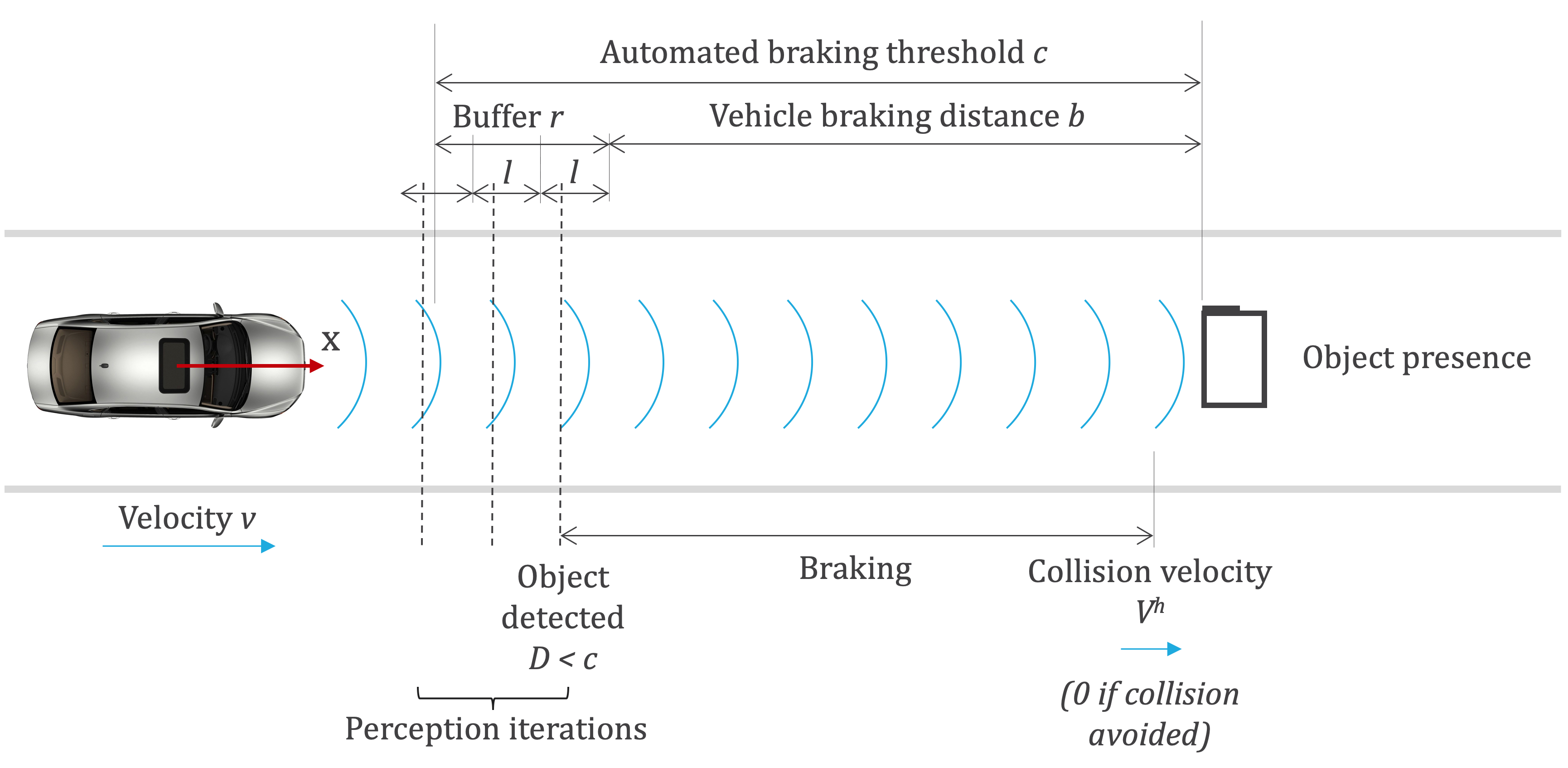}
  \caption{Schematic diagram of the vehicle and the \textsc{odd} setup in the example.}
  \label{fig:schema}
\end{figure}

If the distance estimate to the closest object falls below the braking threshold \gls{threshold}, the car starts braking at the level of maximum surface
utilisation and sustains this level of deceleration until it comes to a full stop. Moreover, in order avoid some mathematical complications, we assume that after each braking event the vehicle is safely restarted (reaccelerated) in such a way that the next stationary object is approached at speed \(\gls{targetvelocity}\) with headway of at least \(\gls{threshold}\). Depending on the nature of false positive detections this could result in the vehicle ending up in a perpetual sequence of braking and reversing. As far as safety is concerned, in the given operating environment, this is not a safety issue. It may make the vehicle functionally useless, but never unsafe. 

The braking functionality works as follows.  The total braking distance to full stop is denoted by \(\gls{brakedist}\), where \(\gls{brakedist} < \gls{threshold}\), with \(\gls{threshold} - \gls{brakedist}\) defined as the 'buffer distance' to a collision. Hence if the car starts braking at a distance no more than \(\gls{threshold}\) and no less than \(\gls{brakedist}\) from an obstacle, then there is no collision; the obstacle is then considered to move from the roadway, and the vehicle may proceed onward. A schema of these distances can be found in Figure~\ref{fig:schema}, where we use the notation \(l:=\gls{targetvelocity}/\gls{frequency}\) for the distance travelled by the vehicle between sequential iterations of the perception system. Also, the frequency of the perception system iterations is such that \(\gls{numestimates} = \lfloor \gls{bufferdist} \gls{frequency}/\gls{targetvelocity} \rfloor \geq 1 \), which is the guaranteed number of distance estimates the perception system generates while traversing the buffer distance. Also, let \(\gls{truedist}_{t}\) be the true distance to the closest object in front of the vehicle at time \(t\) and \(\gls{estdist}_{t}\) the estimate of this distance produced by the perception system. The brakes thus engage at any time point \(t\) when the system determines that \(\gls{estdist}_{t} < c\). 

The autonomous driving system and \acrshort{odd} can now be combined to define a simple, potentially hazardous, driving scenario.  As the vehicle state evolves over time while driving along the defined road section it will sequentially encounter multiple stationary objects, each of which must be avoided through proper activation of the automatic braking functionality.  As the vehicle approaches an object, the intended functionality of the design is to detect the object and activate the braking system and stop the vehicle prior to collision.  A hazardous condition will develop during the scenario if the system fails to properly estimate the distance to the object, and thereby either fails to brake or triggers braking after the minimum safe stopping distance of \(\gls{truedist}_{t} = b\) is violated.  Hazardous events caused by the autonomous driving system correspond to situations where the values between \(\gls{truedist}_{t}\) and \(\gls{estdist}_{t}\) differ significantly enough that the condition of \(\gls{estdist}_{t} <  \gls{threshold}\) is never met over the course of the scenario prior to reaching the condition \(\gls{truedist}_{t} < b\). Once such a situation is generated, a collision cannot be avoided by the system as defined.

\section{Robust probabilistic arguments} \label{S:args}

Having set a quantitative safety target in a previous section, we now proceed to showing how rigorous arguments can be constructed to conclusively answer whether the autonomous vehicle meets (argument 1 in Figure \ref{F:IterProcess}) or does not meet (argument 2 in Figure \ref{F:IterProcess}) the safety target.

\subsection{Sufficient conditions at the component-level for meeting a vehicle-level safety target}
\label{sec:upperbound}

When working with modular designs, focusing on sufficient conditions at the component level in relation to the vehicle-level validation target can be beneficial. A sufficient condition for a component is defined so that if it is met, then the validation target is also met (given the knowledge of the functioning of the rest of the system), e.g., \(\E[H]<\epsilon\). Sufficient conditions can be obtained by making assumptions that the world and the rest of the system behaves worse (no better) than it actually does. Useful sufficient conditions are obtained from a detailed understanding of the \acrshort{odd} including its probabilistic aspects, such as vehicle dynamics, physics of sensing technologies, and the system architecture with its individual components and their dependencies. To justify that sufficient conditions are met at a component level, statistical justifications can be made by utilising practically manageable amounts of data, as the existing knowledge about the problem is taken into account. Focusing on sufficient conditions at a component level can yield not only validation cost reductions but also allow the reuse of much of the safety analysis in case of a change in the ODD or in a system component. Another advantage of such an approach is that arguments for meeting more elaborate quantitative targets, e.g. incorporating fairness considerations \cite{mehrabi_survey_2019}, can be more practically achievable through meeting their sufficient condition targets at a component level. Analogously, when disproving the safety of the system, it is enough to show that necessary safety conditions at component level are not met, which is a focus of section \ref{sec:lowerbound}.

\subsection{Statistical modular safety argument using sufficient conditions}
\label{sec:safetybysufficient}

Consider a single approach of a randomly picked obstacle from the route. In particular, we only consider when the obstacle gets closer than \(\gls{threshold}\) to the vehicle. Any braking happening earlier than this will simply result approaching the obstacle anew according to the system specification. Let \(\gls{hitvelocity}\) denote the speed at which the obstacle is hit with \(\gls{hitvelocity} = 0\) if the vehicle stops appropriately and does not hit
the obstacle. The situation thus far is summarised in Figure~\ref{fig:schema}. Thus in this specific example, we have the decomposition

\begin{equation}\label{eq:fromsingleapproach}
  \E[\gls{numcollisions}] = \P(V_h>0) \E[\gls{numpedestrians}].
\end{equation}

In addition, let us denote \(l_{0} = \gls{threshold}, l_{1}= \gls{brakedist} + \gls{numestimates} \gls{targetvelocity}/\gls{frequency}, l_{2}= \gls{brakedist} + (\gls{numestimates}-1) \gls{targetvelocity}/\gls{frequency}, \ldots , l_{\gls{numestimates}} = \gls{brakedist}+\gls{targetvelocity}/\gls{frequency},  l_{\gls{numestimates}+1} = \gls{brakedist}\).
From the system specification, a depth estimate update is received every
distance \(\gls{targetvelocity}/\gls{frequency}\) travelled so that by construction there exist unique times \(t_{1}, \dotsc, t_{\gls{numestimates}-1}\) such
that \(\gls{truedist}_{t_{j}} \in [l_{j+1}, l_{j})\). Thus \(\gls{truedist}_{t_{1}}, \gls{truedist}_{t_{2}},
\dotsc, \gls{truedist}_{t_{\gls{numestimates}}}\) and \(\gls{estdist}_{t_{1}}, \gls{estdist}_{t_{2}}, \dotsc,
\gls{estdist}_{t_{\gls{numestimates}}}\) correspond to true distances and distance estimates
produced while the obstacle being approached passed through the buffer region. Apart from a possible observation at a distance between \(l_{0}\) and \(l_{1}\) these are all such pairs.
Also, let write \(\gls{esterror}_{j} :=
\gls{estdist}_{t_{j}} - \gls{truedist}_{t_{j}}\) for the corresponding estimation
errors.

While the perception system operates independently on each received frame of
sensor data the sensor data itself is certainly not independent. We therefore
know little about the dependence structure between \(\gls{esterror}_{1}, \dotsc, \gls{esterror}_{\gls{numestimates}}\). Nevertheless, we can certainly say that
\begin{equation} \label{eq:upperboundtrick}
  \begin{split} 
    \P(\gls{hitvelocity} > 0)
    &=
    \P(\text{no brake event in buffer area})  \\
    &\leq
    \P(\text{\(\gls{estdist}_{t_{j}} > \gls{threshold}\) for all \(j = 1, \dotsc, \gls{numestimates}\)}) \\
    &\leq
    \min_{j=1,\dotsc,\gls{numestimates}} \P(\gls{estdist}_{t_{j}} > \gls{threshold}).  
    \end{split}
\end{equation}
The first inequality is an equality unless there is an observation at a distance between \(l_{0}\) and \(l_{1}\) of the obstacle and the second inequality is an equality in case of the worst
case dependence between estimation errors. This would be an «all or nothing»
scenario where all \(\gls{estdist}_{t_1}, \dotsc, \gls{estdist}_{t_{\gls{numestimates}-1}}\) are greater
than \(\gls{threshold}\) as soon as this is the case for the one with lowest marginal
probability of being greater than \(\gls{threshold}\).

Thus in this specific example, we get that the risk
\begin{equation*}
    \E[\gls{numcollisions}]
    \leq
    \left( \min_{j=1,\dotsc,\gls{numestimates}} \P(\gls{estdist}_{t_{j}} > \gls{threshold} ) \right)  \E[\gls{numpedestrians}].
\end{equation*}
Each of \(\P(\gls{estdist}_{t_{j}} > \gls{threshold})\) for
\(j=1,\dotsc,\gls{numestimates}\) depends only on the performance of sensors
while \(\E[\gls{numpedestrians}]\) is purely a property of the environment. Both
can be studied statistically and controlled by, for example, estimating confidence intervals. The above analysis then transfers
results of such a statistical analysis to a potential safety argument at the
vehicle-level.

\begin{example} (Modular safety argument via sufficient conditions) \label{e:numerical}
To illustrate the argument above even more concretely,  let us assume  that the validation target is less than 1 collision in 100000 km of driving on average and to be proven with a confidence level of at least  \(1-\alpha\), i.e. \(\E[\text{Collisions per km}]< 10^{-5}\) with confidence at least \(1-\alpha\). Suppose that one estimates that there is less than 1 stationary road object per 100 km (e.g. from traffic or driving statistics) with confidence at least \(1-\alpha_1\), i.e. \(\E[\text{Objects per km}]< 0.01\) with confidence at least \(1-\alpha_1\). In addition, suppose that one also estimates that the detection performance \(\P\left(D_{t_{\gls{numestimates}}}>c\right)<0.001\) with confidence at least \(1-\alpha_2\). Then, using the proposed \eqref{eq:upperboundtrick} above,  \(\E[\text{Collisions per km}]<0.01 \times 0.001\) with confidence at least \(1-(\alpha_1+\alpha_2)\). If \(\alpha_{1}\) and \(\alpha_{2}\) are chosen such that \(\alpha_{1} + \alpha_{2} = \alpha\) one has then achieved the target. Here the confidence levels are combined as justified by elementary probability rules\footnote{If for any two events \(A\) and \(B\), we have \(\P(A) \geq 1-\alpha_1\) and \(\P(B)\geq 1-\alpha_2\) , then \(\P(A \cap B) = P(\Omega) -\P(\Omega \setminus (A \cap B)) = 1 -  \P((\Omega \setminus A) \cup (\Omega \setminus B)) \geq 1 - (\P(\Omega \setminus A) + \P(\Omega \setminus B)) \geq 1 - (\alpha_1+\alpha_2)\).}. 

If the two confidence intervals are independent the requirement directly weakens to \(\alpha_{1} + \alpha_{2} - \alpha_{1}\alpha_{2} < \alpha\). This means weaker individual confidence levels may be used, resulting in smaller required sample sizes and, in the end, lower testing costs. Such an assumption of independence would be reasonable if the intervals were constructed based on independent data sets, as may be realistic in case one is based on general road statistics and the other on actual test drives.

Figure~\ref{fig:gsn} contains a diagrammatic version of such a modular argument in the GSN-notation \cite{gsn}, elucidating the argument structure potentially relevant for arguing safety of some other autonomous driving functions, not just autonomous braking.

\end{example}

\subsection{Numerical example of sample size estimation}
\label{numeric}

\begin{figure}[t]
\begin{subfigure}[b]{0.5\textwidth}
  \centering{}
  \includegraphics[width=\textwidth]{{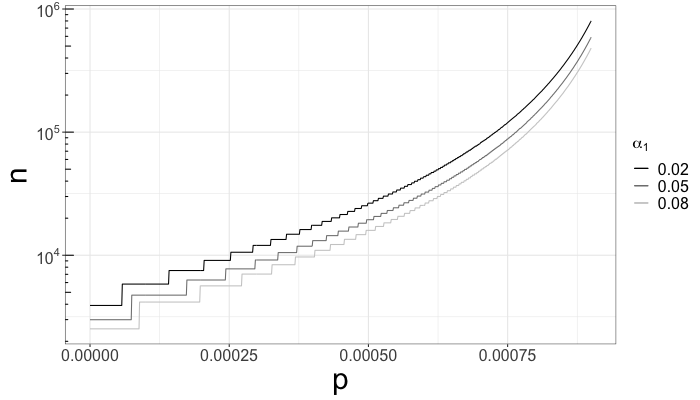}}
  \caption{\(p_c = 0.001\) and \(\alpha =0.1\)}
  \label{fig:p1}
  \centering{}
\end{subfigure}%
  \begin{subfigure}[b]{0.5\textwidth}
    \centering{}
  \includegraphics[width=\textwidth]{{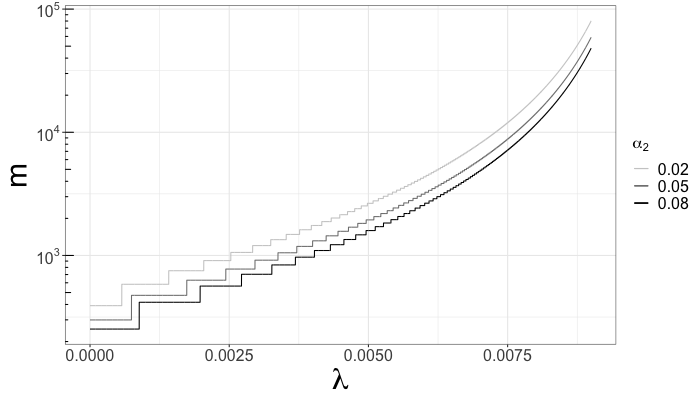}}
  \caption{\(\lambda_c = 0.01\) and \(\alpha =0.1\)}
  \label{fig:lambda1}
  \end{subfigure}
  
\begin{subfigure}[b]{0.5\textwidth}
  \centering{}
  \includegraphics[width=\textwidth]{{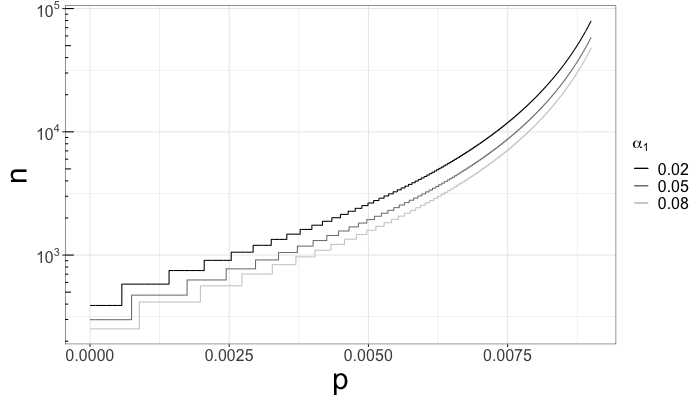}}
  \caption{\(p_c = 0.01\) and \(\alpha =0.1\)}
  \label{fig:p2}
\end{subfigure}  
  \begin{subfigure}[b]{0.5\textwidth}
    \centering{}
  \includegraphics[width=\textwidth]{{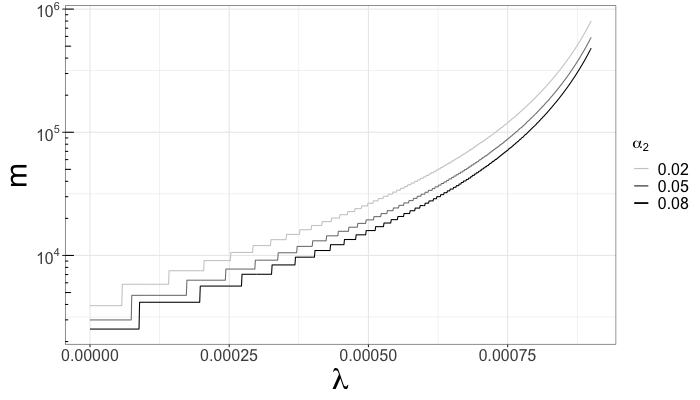}}
  \caption{\(\lambda_c = 0.001\) and \(\alpha =0.1\)}
  \label{fig:lambda2}
  \centering{}
\end{subfigure}
\caption{Required sample sizes to achieve a probability of 80\% of successfully
  establishing bounds at different underlying values of distance over-estimation
  and obstacle intensities given different target values and
  significance levels requirements. Differently shaded lines correspond to decomposing \(0.1 = \alpha = \alpha_{1} + \alpha_{2}\) as \(0.02 + 0.08\), \(0.05 + 0.05\), and \(0.08 + 0.02\). \label{fig:pandlambda}
}
\end{figure}

To get an idea of what amount of data would be required to perform an analysis
as in Example~\ref{e:numerical} let \(p=\P\left(D_{t_{\gls{numestimates}}}>c\right)\) be the
underlying probability of miss-classification and \(\lambda\) the underlying
intensity of obstacles. Suppose we want to prove that \(p<p_c\) and \(\lambda <
\lambda_c\) for some specified \(p_c\) and \(\lambda_c\) such that \(\lambda_c
p_c\leq \epsilon\) with confidence at least \(1-\alpha\).
Example~\ref{e:numerical} would correspond to \(p_{c} = 0.001\) and \(\lambda_{c} =
0.01\).

Using exact one-sided binomial and Poisson confidence intervals (it is possible to use approximate confidence intervals, but some additional care has to be taken to ensure that the choice of confidence intervals is conservative \cite{tony_cai_one-sided_2005}) one may attempt this using two data sets, one of \(n\) samples
of potential perception system failures and one given by observing the number of
obstacles in \(m\) kilometres of road segment. For each actual value of \(p\)
and \(\lambda\), choices of \(n\) and \(m\), and chosen confidence levels
\(\alpha_{1}\) and \(\alpha_{2}\) for the two confidence intervals one then gets
probabilities \(W_p(p)\) and \(W_{\lambda}(\lambda)\) of successfully establishing the
required bounds. These numbers are exactly the power levels of the hypothesis
tests built from these confidence intervals for the alternative hypotheses \(p\)
and \(\lambda\).

Figure~\ref{fig:pandlambda} illustrates the required samples sizes for number of
such scenarios. In Figure~\ref{fig:p1} we set \(p_c=0.001\) and the plot shows
for each \(p < 0.9p_c\) (\(x\) axis) the smallest number \(n\) (\(y\) axis) of
sample size needed to have a power of at least 0.8 at \(p\) and significance
levels \(\alpha_{1} = 0.02, 0.05, 0.08\).

In these plots the same line types for, say, \(\alpha_{1}\) correspond to line
types for \(\alpha_2=\alpha-\alpha_1\). So a smaller confidence level in one
parameter could be balanced by a higher confidence level in another parameter.
Using system and domain expert judgement on feasible underlying values of
\(p\) and \(\lambda\), one can then optimise the likelihood of a successful
outcome of the safety argumentation with an optimally small amount of data.

In Figure~\ref{fig:lambda1}, we set \(\lambda_{c}=0.01\) and show for each
\(\lambda < 0.9\lambda_c\) the smallest number \(m\) of kilometres of data
needed to have power 0.8 at \(\lambda\) and significance \(\alpha_{2} = 0.02,
0.05, 0.08\).

A similar situation is illustrated in Figures~\ref{fig:p2}~and~\ref{fig:lambda2}
where we set \(p_c=0.01\) and \(\lambda_c=0.001\). This would correspond to a
weaker perception system with rarer obstacle events.

We may also look at Figure ~\ref{fig:p1}~and~Figure~\ref{fig:lambda2} to analyse the case \(\epsilon = 10^{-6}\) and \(\alpha=0.1\) as \(p_c=10^{-3}\) and \(\lambda_c=10^{-3}\). If we have a priori estimates that \(\lambda=0.0005\) and \(p=0.0005\) then to have tests with significance levels \(\alpha_1=0.08\), \(\alpha_2=0.02\) and \(W_{p}(0.0005), W_{\lambda}(0.0005)>0.8\) we would need \(n\) to be 15922 and \(m\) to be 26497.6 km. More sample size values can be seen in Table \ref{table:1}. These sample sizes appear to be practically manageable. Also, according according to NHTSA \cite{nhtsa_report_2017}, in USA there is less than 1 injury reported per \(10^{6}\) vehicle miles driven, so this example would give a safety bound that is close to the real world observed injury rate. But of course, this example depends on the actual parameter value of obstacle density \(\lambda\) and the selected confidence level \(1-\alpha\) might be needed to be larger. Additional required sample size calculations for different parameter values are provided in Figure \ref{fig:extra_graphs} in the appendix.

\begin{table}
\centering
\begin{tabular}{lcc} 

 \toprule
  \(\alpha_i\) & \(n\) & \(m\) \\ 
  \midrule
0.08 & 15922 & 15924.71 \\
0.05 & 19439 & 19442.58 \\
0.04 & 21181 & 21184.97 \\
0.03 & 23076 & 23079.97 \\
0.025 & 24736 & 24740.22 \\
0.02 & 26493 & 26497.63 \\
0.01 & 31839 & 31845.37 \\
0.005 & 35939 & 35946.28 \\
 \bottomrule
 \end{tabular}
 
 \caption{Values for \(n\) and \(m\) to have tests of whether \(p\) and \(\lambda\) are smaller than \(0.001\) with significance at least \(1-\alpha_i\) and power levels at least 0.8 at \(0.0005\).}
\label{table:1}
\end{table}

\subsection{Disproving vehicle-level safety through not meeting necessary
  component-level conditions}
\label{sec:lowerbound}

In some circumstances one will want to perform converse analysis to the above, that is, to prove that the system is, conclusively,
unsafe. This may, for example, serve to show that the issue is actually with the
system, and not with the conservative worst-case analysis performed when
attempting to make the safety argumentation.

As in Section~\ref{sec:safetybysufficient}, we may
reduce the problem to studying the approach of a single randomly picked obstacle. In the ideal case \(\gls{bufferdist}=\gls{numestimates} v/f\) so that the perception system generates exactly \(\gls{numestimates}\) distance updates when traversing the intended braking window of length \(\gls{bufferdist}\). Otherwise assuming \(\gls{numestimates}+1\) rather than \(\gls{numestimates}\) opportunities for detection is certainly non-pessimistic. Therefore let \(Z_{0}\) is the estimation error of an additional detection opportunity within the first interval \([l_1, l_0)\).

When it comes to dependency of the estimation errors \(\gls{esterror}_{0}, \dotsc \gls{esterror}_{\gls{numestimates}}\) one would expect at least for values adjacent in time, that over-estimating the real distance be positively correlated. That is to say, over estimating in one frame would make it more likely to over estimate also on the next. In any case, without strong arguments against such positive dependence the system should be shown to be safe assuming a positive dependence between over-estimations. The most optimistic situation one can then wish is that \(\gls{esterror}_{0}, \dotsc, \gls{esterror}_{\gls{numestimates}}\) are independent. As already argued the
requirements on rigour can be lowered when arguing \emph{against} the
safety of the system\footnote{If one thinks about the theoretically optimal dependence structure for safety, it is the situation where \(\gls{esterror}_{0}, \ldots \gls{esterror}_{\gls{numestimates}}\) are such that exactly one, or no, correct distance estimate occurs. Intuitively, only one correct detection during the
approach is required and in such situation no correct detections are wasted. In this analysis equation~\ref{eq:optimistic} turns into a sum of the marginal probabilities.}. In case of independence of errors, we would have
\begin{equation} \label{eq:optimistic}
  \begin{split}
    \P(\gls{hitvelocity} > 0)
    &\geq
    \prod_{j = 0, \dotsc, \gls{numestimates}} \P(\gls{esterror}_{j} > \gls{threshold} - \gls{truedist}_{t_{j}}) \\
    &\geq
    \left(\min_{j = 0, \dotsc, \gls{numestimates}} \P(\gls{esterror}_{j} > \gls{threshold} - \gls{truedist}_{t_{j}}) \right)^{\gls{numestimates}+1}.
  \end{split}
\end{equation}

For some perception systems, it might be justified from the system design
knowledge that overestimation errors are decreasing with distance, in
particular, implying that \(\P(\gls{esterror}_{\gls{numestimates}} \geq z) \leq
\P(\gls{esterror}_{j} \geq z)\) for all \(z > 0\) and \(j \leq \gls{numestimates}\). Then
\begin{equation*} 
  \P(\gls{hitvelocity} > 0)
  \geq
  \P(\gls{esterror}_{\gls{numestimates}} > \gls{threshold} - \gls{truedist}_{t_{\gls{numestimates}}})^{\gls{numestimates}+1}.
\end{equation*}

As in Section~\ref{sec:safetybysufficient}, all relevant quantities can now be estimated with
confidence intervals from the perception system performance data and data about
the environment. More precisely
\begin{equation*}
  \begin{split}
    \E[\gls{numcollisions}]
    &=
    \P(\gls{hitvelocity} > 0) \E[\gls{numpedestrians}] \\
    &\geq
    \P(\gls{esterror}_{\gls{numestimates}} > \gls{threshold} - \gls{truedist}_{t_{\gls{numestimates}}})^{\gls{numestimates}+1}
    \E[\gls{numpedestrians}].
  \end{split}
\end{equation*}
This lower bound can be utilised to show that the vehicle-level quantitative
safety target is not met (a similar argument as in Example \ref{e:numerical} can be followed just with inequalities and one-sided confidence intervals reversed). 

\subsection{Statistical estimation considerations}
\label{sec:stat}

The fundamental quantities to control in the two previous sections are
\(\P(\gls{estdist}_{j} > c) = \P(\gls{esterror}_{j} > \gls{threshold} -
\gls{truedist}_{t_{j}})\) for \(j=1,\dotsc,\gls{numestimates}\) and
\(\E[\gls{numpedestrians}]\). In particular, for the bound in
Section~\ref{sec:upperbound} requires controlling
\(\min_{j=1,\dotsc,\gls{numestimates}} \P(\gls{estdist}_{j} >
\gls{threshold})\).

Unless one is able to make a case that a monotonicity assumption such as in
Section~\ref{sec:lowerbound} one will not know at which \(j\) the minimum is
achieved. It would therefore generally not be enough to control
\(\P(\gls{estdist}_{j} > \gls{threshold})\) separately for each \(j\) and take
the minimum. Each separate estimate adds another level of uncertainty.
A simple way of compensating for this is to observe that a minimum is certainly
bounded by an average, so that for any probability vector \(p_{1}, \dotsc,
p_{\gls{numestimates}}\) we have
\begin{equation}\label{eq:minmean}
  \min_{j=1,\dotsc,\gls{numestimates}} \P(\gls{estdist}_{t_{j}} > \gls{threshold})
  \leq
  \sum_{j=1}^{\gls{numestimates}}p_{j}\P(\gls{estdist}_{t_{j}} > \gls{threshold})
  =
  \P(\gls{estdist}_{t_{J}} > \gls{threshold}),
\end{equation}
where \(J\) is independent of \(\gls{estdist}\) and \(\P(J = j) =
p_{j}\). In particular, if \(p_{1} = 0, \dotsc, p_{\gls{numestimates}-1} = 0, p_{\gls{numestimates}} = 1\), then
\(\gls{estdist}_{t_{J}} = \gls{estdist}_{t_{\gls{numestimates}}}\). More generally if we randomly
sample frames from the perception system in use, we may choose \(J\)
according to the sampling distribution. This works even if this distribution is
unknown, as long as \(J\) remains independent of \(\gls{estdist}\). In
practical terms, this excludes strategies where sampling is based on information
gathered from the sensors, such as observing the system when an obstacle is
estimated to be within a certain distance.

The quantity
\(\P(\gls{estdist}_{t_{J}} > \gls{threshold})\) can be estimated from
individual frames if we sample 
uniformly within each interval \([l_{j+1}, l_{j})\).
The problem of estimating is  \(\P(\gls{estdist}_{t_{J}} >
\gls{threshold})\) is that of estimating the probability parameter of the Bernoulli distribution. However, note that to provide a safety argument as in Example \ref{e:numerical}, we need to use a conservative one-sided interval estimator (c.f.\  \cite{tony_cai_one-sided_2005}).  

In order to control the term
\(\E(\gls{numpedestrians})\) we assume we have observed a number of road
segments with accurate counts of the number of obstacles. This data need not be
from the same source as the perception system performance data. Indeed, the
perception system data may be produced by designed laboratory experiments while
the data on number of road segment obstacles may be derived from publicly
available data sources covering the geographical area designated by the
\acrshort{odd}.

Some care should be taken unless the observed road
segments are all of equal length \(\gls{drivedist}\,\si{\kilo\metre}\). Longer
segments should result in higher variance in the number of obstacles, simply
because there is more road segment for the total number to vary on.  A reasonably robust methodology (while not
obviously formally conservative) would be to
consider the number on each road segment to follow a Poisson distribution with
mean proportional to the length. A one-sided confidence interval for the expected number
for a \(\gls{drivedist}\,\si{\kilo\metre}\) road segment can be then estimated
using known statistical methods as the resulting model is a generalised linear model
(Poisson-regression) \cite{greene_econometric_2003}.

\section{Refining the model and analysis} \label{S:refining}

The authors recognise that both the model and the analysis given here is greatly simplified compared to a model of a real autonomous driving system. This is intentional, both for reasons of pedagogy and because of the size necessary for a safety analysis even approaching one as might be necessary to certify a system for deployment. Nevertheless, we believe it highlights some important and central issues that such an argumentation would probably have to include. That being said, it is still worth pointing out some of the directions in which one would have to extend the proposed approach to apply to realistic systems and thus give at least outline of what a complete argumentation strategy would look like.

As already noted, in the current setup false positive braking events pose no safety hazard, appearing simply as, presumably, an annoyance to the driver. In reality this is of course not the case. An unnecessarily triggered automated braking event can cause both direct harm to the driver and vehicle passengers, and be the cause of a collision with a following vehicle, triggering the onset of a serious accident. Extending the analysis to cover this would obviously necessitate analysing the probabilities of false positive braking events. In case one knows, or suspects, specific triggering conditions for false positives the situation is not too different. One would expand the model enough to be able to express these triggering conditions, and proceed similarly to the analysis here: create statistical bounds for the frequency with which the conditions occur and the probability of, now, a false positive in the presence of these triggering conditions. Most real deployment environments would be significantly more complex than modelled here and one might expect hazardous behaviour of the system to occur primarily due to complex and rare co-occurrences. In some cases these might be predicable by extrapolating from rare events in existing data, so that one might identify additional potential triggering conditions. For systems whose principles of functioning are sufficiently transparent and understood, this may result in enough coverage to argument for the safety of the overall system. For black-box systems, a «perfect storm» of unfortunate conditions might simply be impossible to prepare for by means other than brute force testing. 
Another dimension of making such an extension would be to quantify the expected harm from such unnecessary braking events. As a proposal, one could separately model the distance to other vehicles in traffic, their braking behaviour, and the resulting concrete harm of an actual collision as a function of impact velocities. Each one, again, would be made with an eye towards reasonable worst case scenarios. Some of this analysis would mirror the way the distribution of obstacles is quantified here, while others can be based on published empirical evidence of injuries resulting from accidents (e.g.\ literature reviews such as \cite{leaf1999literature}). 

Going beyond the simple example analysed, further modular decomposition would be useful for more realistic autonomous driving systems typically relying on multiple sensors and incorporating additional tracking and trajectory prediction components. This introduces additional complex dependencies between the stochastic quantities analysed. 
One may, for example, have the option of analysing the perception system as a composition of two components, one tracking dynamic objects in the environment (over time) and one that is tasked with detecting objects. Performance of the former would have to be analysed in terms of its ability to track objects while in range of the vehicle, while the latter could be analysed analogously to what has been presented in this paper. Moreover, dependence between sensors can be simplified by physical models. Light determines the signal of any camera, but should not change the signal produced by a LiDAR module. Conditionally on factors such as atmospheric conditions and the actual macroscopic configuration of the scene, any additional noise components disturbing either should exhibit at most very weak dependence.

\begin{figure}[h]
  \centering{}
  \includegraphics[width=\textwidth]{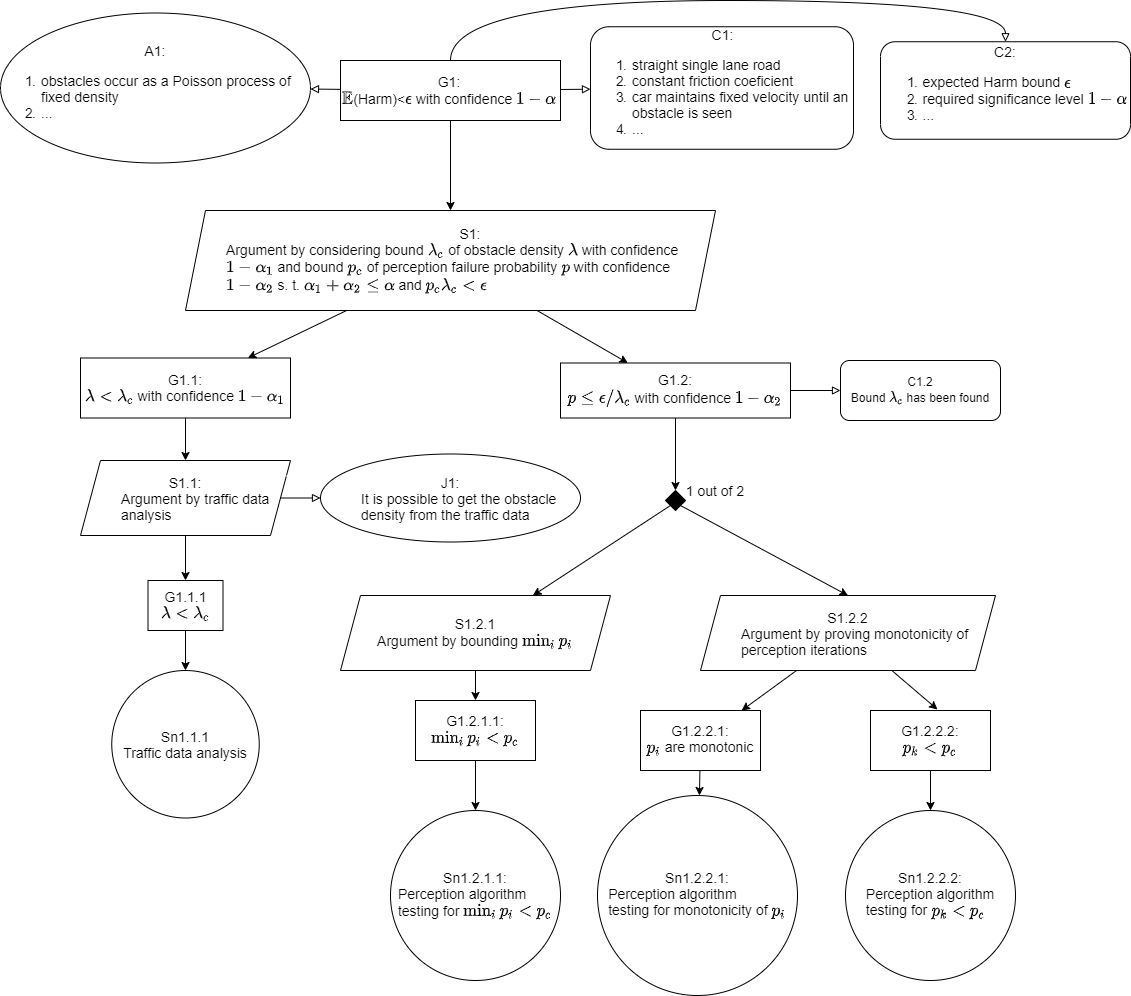}
  \caption{Diagrammatic form in GSN\cite{gsn} of the main part of the argument outlined in the paper.}
  \label{fig:gsn}
\end{figure}

When making safety argumentation for other automated driving functions, parts of the analysis could be potentially reused. When analysing something like an adaptive cruise control function involving a camera component, the bounds established for distance over-estimation might be possible to reuse. This would correspond to copying over the sub-tree rooted at G1.2 in the GSN diagram in Figure~\ref{fig:gsn}. Similarly, the sub-tree rooted at G1.1, could be appropriately reused in situations where harm is related to obstacles on the road.

Though the study of other automated driving functions including lateral control like `lane keep' is outside the scope of this paper and is a topic for a separate publication, the general principles of propagating sufficient and necessary conditions to component level to allow for a modular analysis are still applicable when analysing the safety of these other functions. For such modular safety analyses, the relevant aspect of driveable surface identification, path planning, control, and actuation components would need to be included. 

\section{Conclusions}

The probabilistic and statistical modelling approach presented in this article enables the rigorous and quantitative analysis and argument for the level of safety generated by the operation of an autonomous driving function. It is achieved  through analysis of sufficient and necessary conditions for the constituent components and exploiting available probabilistic decompositions. Besides the clear benefits of facilitating modular design, this approach enables proving that sufficient safety conditions are met at the component level using data sets of reduced size and so smaller cost in comparison to the amounts required for validation by vehicle-level road testing. In the simple example studied, sufficient and necessary conditions as well as statistical estimation techniques discussed can be refined further, however, the aim of the paper here was to elucidate the main principles and benefits of rigorous statistical modular safety argumentation for autonomous driving rather diving deep into mathematical and statistical technicalities.

\section*{Acknowledgements}

We would like to thank the anonymous reviewers for constructive comments. Our thanks go also to Mark Costin, PhD, at NVIDIA for useful discussions and practical feedback.

\printbibliography{}

\appendix
\section{Supplementary material}

\begin{figure}[h] 
\begin{subfigure}[b]{0.5\textwidth}
  \centering{}
  \includegraphics[width=\textwidth]{{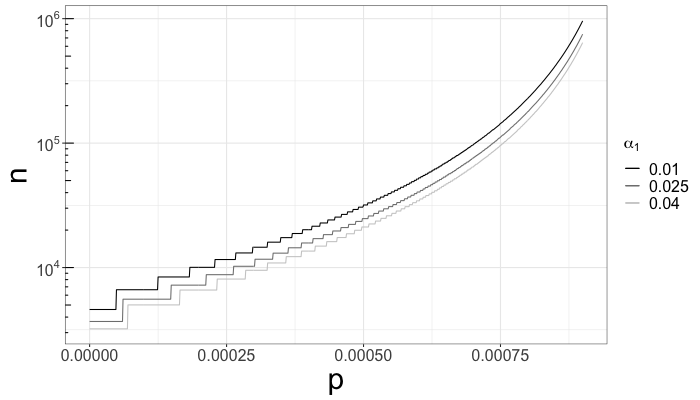}}
  \caption{\(p_c =0.001\) and \(\alpha = 0.05\)}
  \centering{}
\end{subfigure}%
  \begin{subfigure}[b]{0.5\textwidth}
    \centering{}
  \includegraphics[width=\textwidth]{{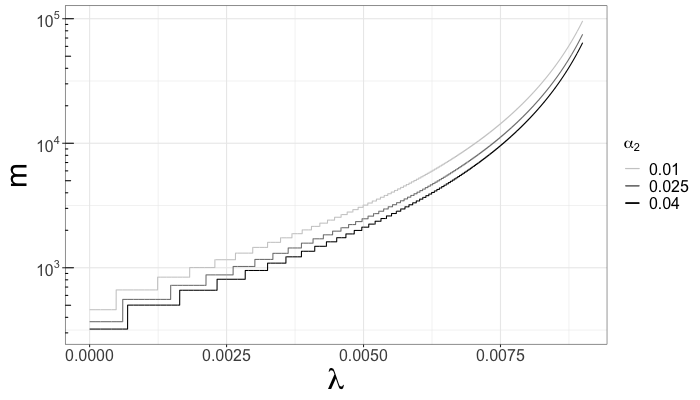}}
  \caption{\(\lambda_c = 0.01\) and \(\alpha = 0.05\)}
  \end{subfigure}
  
\begin{subfigure}[b]{0.5\textwidth}
  \centering{}
  \includegraphics[width=\textwidth]{{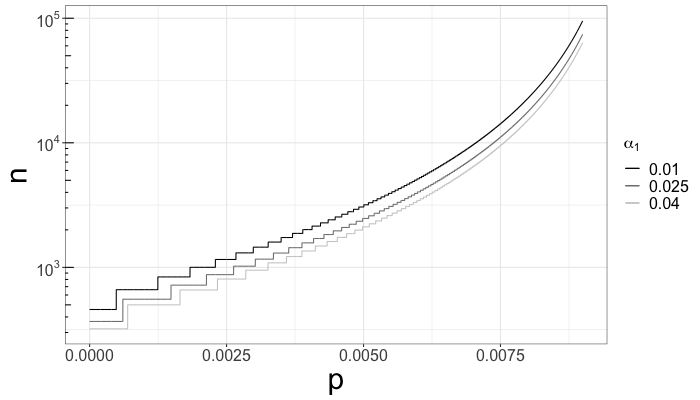}}
  \caption{\(p_c = 0.01\) and \(\alpha = 0.05\)}
\end{subfigure}  
  \begin{subfigure}[b]{0.5\textwidth}
    \centering{}
  \includegraphics[width=\textwidth]{{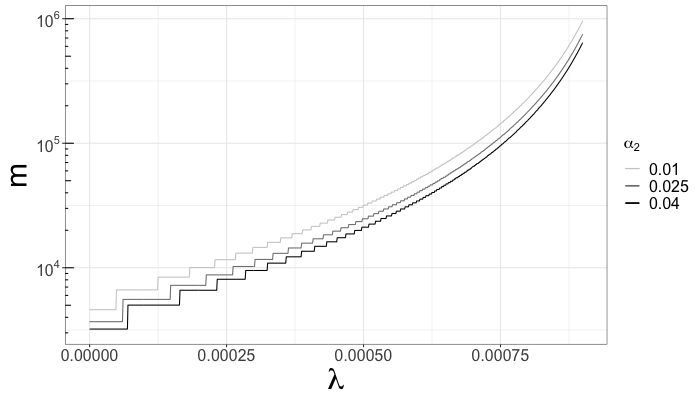}}
  \caption{\(\lambda_c = 0.001\) and \(\alpha = 0.05\)}
  \centering{}
\end{subfigure}

\begin{subfigure}[b]{0.5\textwidth}
  \centering{}
  \includegraphics[width=\textwidth]{{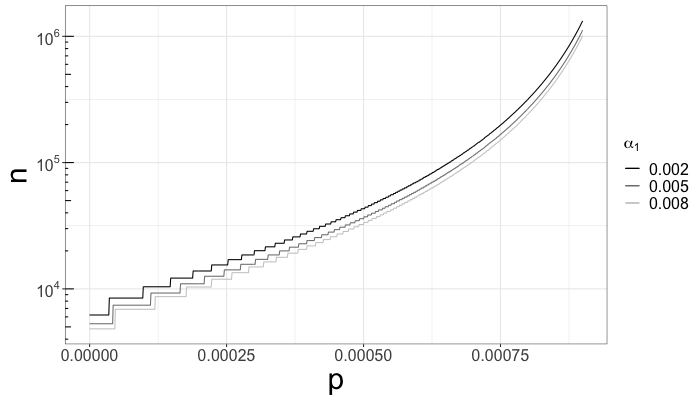}}
  \caption{\(p_c = 0.001\) and \(\alpha = 0.01\)}
  \centering{}
\end{subfigure}%
  \begin{subfigure}[b]{0.5\textwidth}
    \centering{}
  \includegraphics[width=\textwidth]{{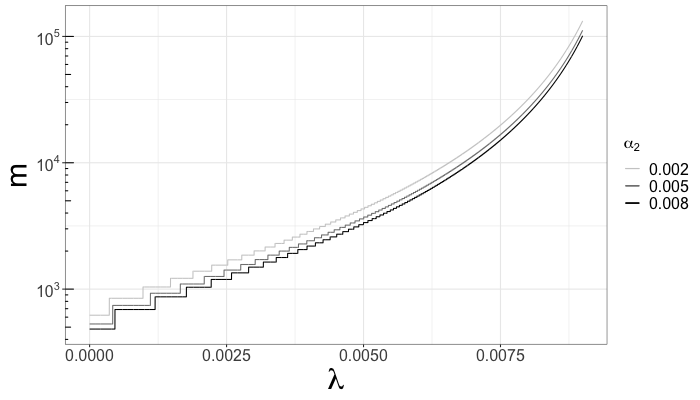}}
  \caption{\(\lambda_c = 0.01\) and \(\alpha = 0.01\)}
  \end{subfigure}
  
\begin{subfigure}[b]{0.5\textwidth}
  \centering{}
  \includegraphics[width=\textwidth]{{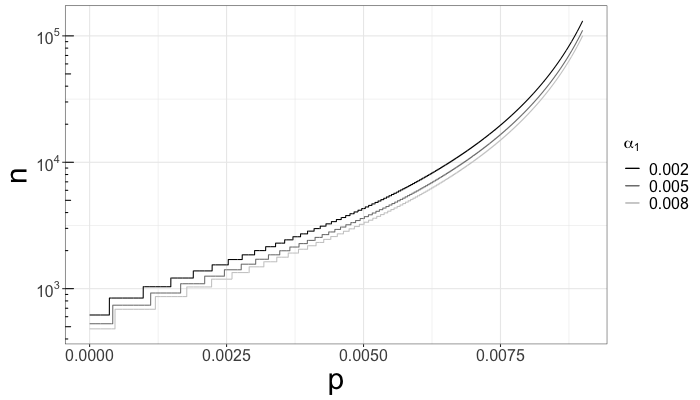}}
  \caption{\(p_c = 0.01\) and \(\alpha = 0.001\)}
\end{subfigure}  
  \begin{subfigure}[b]{0.5\textwidth}
    \centering{}
  \includegraphics[width=\textwidth]{{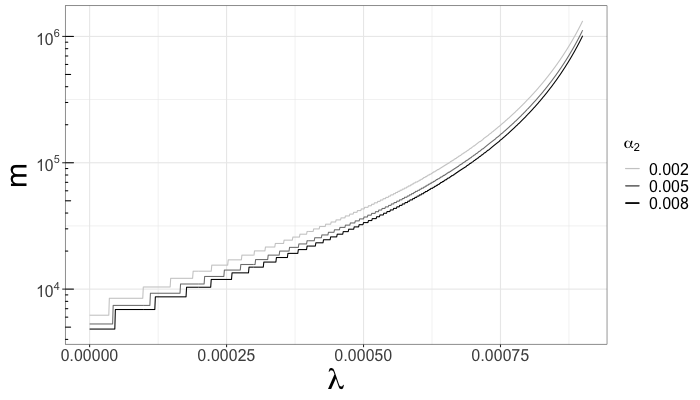}}
  \caption{\(\lambda_c = 0.001\) and \(\alpha = 0.01\)}
  \centering{}
\end{subfigure}
\caption{Required sample sizes to achieve a probability of 80\% of successfully
  establishing bounds at different underlying values of distance over-estimation
  and obstacle intensities given different target values and
  significance levels requirements.} \label{fig:extra_graphs}
\end{figure}

\end{document}